\documentclass[epj,referee]{svjour}
\usepackage{graphics}
\usepackage{amssymb}
\usepackage{amsmath}
\begin{document}
\title{Optical Diodes and Non-Reciprocal Wave Propagation}
\subtitle{ The one-way light-speed anisotropy in  parity-odd linear, and  non-linear crystals}
\author{Qasem Exirifard\inst{1} 
}                     
%
%
\institute{School of Physics, Institute for Research in Fundamental Sciences, Tehran 19538-33511,  Iran
}

%
\date{Received: date / Revised version: date}
%
\abstract{
We report that a triangular Fabry-Perot resonator  filled with a parity-odd linear anisotropic medium  exhibiting the one-way light speed anisotropy acts as a perfect diode.   A Linear crystal such as the nematic liquid crystals whose molecular structures break parity  can exhibit the one-way light speed anisotropy. The one-way light speed anisotropy also can  be induced in a  non-linear medium in the presence of constant electric and magnetic field strengths. 
} 

\maketitle
\section{Introduction}
\label{intro}
Electrical diodes and transistors constitute  the two pillars of the electronic logical computations. The logical computation, however, can be performed over any kind of currents provided that the counterparts of diodes and transistors  are known for that current. Perhaps the photon's current is a good substitute for the electronic currents. So it is compelling  to construct optical diodes and transistors. 

The optical  diodes or isolators (non-reciprocal wave propagation ) have been constructed based on  quantum hall effects \cite{NatureDiode}, nonreciprocal wave guides \cite{ScienceDiode}, non-linear optics   \cite{NonLinear1,NonLinear2,N2} and various meta-materials \cite{M1,M2,M3,M4,M5} . In this Letter we report  that perfect optical diodes can be constructed based on the one-way light speed anisotropy.  We utilize the  SME electrodynamics to address the light propagation in an anisotropic medium.  This suggests that light propagation in an anisotropic medium generally possesses 19 parameters, 8 out of which are parity-odd. These parity odd parameters contribute to the one-way light speed anisotropy. The one-way light speed anisotropy can occur in linear crystals whose molecular structure violates parity. This includes the nematic liquid crystals. The experimental optical community, however so far,  has not been interested in measuring the one-way light speed anisotropy. We urge the community to measure the one way light speed anisotropy in  anisotropic materials,  in particular in the nematic liquid crystals. 

The Letter is organized as follows. First we review the light propagation in a general anisotropic media.  We use the language of the mSME. We show that a material with a frequency dispersion relation generally has eight parameters contributing to the one-way light speed anisotropy. These parameters can not be set to zero from outset when the molecular structure of the material violates the parity. For example no theoretical argument can be employed to argue that all these eight parameters are vanishing in the nematic liquid crystals. We also note that turning on a constant field strength can induce the one-way light speed anisotropy in   non-linear optical media.  

In the third section, we show how to construct optical diode based on the one-way light speed anisotropy: intrinsic anisotropy or the induced one in a non-linear material.   In the fourth section we ask the experimental groups to empirically measure the one-way light speed anisotropy in various media. At the end we provide the summary and conclusions.

\section{Light Propagation in a general medium}
In this section, we review and utilize  the language of the Standard Model Extension \cite{Kostelecky:2008ts} to address the light propagation in a given material. The SME electrodynamics addresses the break of the Lorentz symmetry in the vacuum. So it can naturally be used to address the break of Lorentz symmetry or isotropy in a medium.  
\subsection{Electromagnetism of  a linear anisotropic medium}
The electrodynamics in the vacuum is governed by the following Lagrangian density:
\begin{equation}
{\cal L}\,=\, -\frac{1}{4} F_{\mu\nu} F^{\mu\nu}
\end{equation}
This Lagrangian density should be modified in the presence of matter. The light propagation in a linear anisotropic medium  is described by
\begin{equation}\label{MGEh}
{\cal L} \,=\, -\frac{1}{4} F_{\mu\nu} F^{\mu\nu}\,-\frac{1}{4} (k_{F})_{\mu\nu\lambda\eta} F^{\lambda\eta} F^{\mu\nu},
\end{equation}
where $k_F$ is a tensor. Its components are dimensionless.  Note that the above Lagrangian describes light propagation in a CPT-even material with frequency dispersion relation.  The materials with spatial dispersion relation must be described by the non-minimal SME models. CPT-odd terms need dimensionfull parameters, so they contribute to the spatial dispersion relation.  In this letter we just consider CPT-even materials with frequency dispersion relations.  
\eqref{MGEh} encodes the optical property of the media. $(k_{F})_{\mu\nu\lambda\eta}$ has the symmetries of the Riemann tensor, so only $20$ out of its $256$ components are algebraically independent. Its double trace should be zero.   So only $19$ algebraically independent components of the $(k_{F})_{\mu\nu\lambda\eta}$ contribute to the equations of motion of the gauge field. These components can be enclosed in the parity-even and parity-odd subsectors, respectively represented by matrices $\tilde{k}_e$ and $\tilde{k}_o$:
\begin{eqnarray}
\left(  \widetilde{\kappa}_{e+}\right)  ^{jk} &  =& \frac{1}{2}(\kappa_{DE}+\kappa_{HB})^{jk},\\
 \kappa_{\text{tr}}&=&\frac{1}{3}\text{tr}(\kappa_{DE}),\\
\left(  \widetilde{\kappa}_{e-}\right)  ^{jk} &=&\frac{1}{2}(\kappa_{DE}-\kappa_{HB})^{jk}-\frac{1}{3}\delta^{jk}(\kappa_{DE})^{ii},\\
\left(  \widetilde{\kappa}_{o+}\right)  ^{jk} &  =&\frac{1}{2}(\kappa_{DB}+\kappa_{HE})^{jk},\\
\left(  \widetilde{\kappa}_{o-}\right)^{jk} &=&\frac{1}{2}(\kappa_{DB}-\kappa_{HE})^{jk}~~.
\end{eqnarray}
The $3\times3$ matrices $\kappa_{DE},\kappa_{HB},\kappa_{DB},\kappa_{HE} $ are
given by:
\begin{eqnarray}
\left(  \kappa_{DE}\right)  ^{jk} &  =&-2(k_F)^{0j0k},\\
\left(  \kappa_{HB}\right)  ^{jk}&=&\frac{1}{2}\epsilon^{jpq}\epsilon^{klm}(k_F)^{pqlm}\\
\left(  \kappa_{DB}\right)  ^{jk} &  =-&\left(  \kappa_{HE}\right)^{kj}=\epsilon^{kpq}(k_F)^{0jpq}.\label{P2}%
\end{eqnarray}
where $\widetilde{\kappa}_{e+},\widetilde{\kappa}_{e-},\widetilde{\kappa}_{o-}$ are  traceless and symmetric $3\times 3$ matrices while $\widetilde{\kappa}_{o+}$ is an anti-symetric matrix. $\kappa_{\text{tr}}$ is a number, it represents the isometric LIV term. $\kappa_{\text{tr}}$ represents the refractive index of the isotropic media. In term of this parametrization, the Lagrangian density reads
\begin{eqnarray}
\mathcal{L}&=&
\frac{1}{2}\left[  \left(  1+\kappa_{\text{tr}}\right)
\mathbf{E}^{2}-\left(  1-\kappa_{\text{tr}}\right)  \mathbf{B}^{2}\right] \nonumber\\
&&+\frac{1}{2}\mathbf{E}\cdot\left(  \widetilde{\kappa}_{e+}+\widetilde{\kappa
}_{e-}\right)  \cdot\mathbf{E}
  -\frac{1}{2}\mathbf{B}\cdot\left(  \widetilde{\kappa}_{e+}-\widetilde
{\kappa}_{e-}\right)  \cdot\mathbf{B}\nonumber 
\\&&+\mathbf{E}\cdot\left(  \widetilde
{\kappa}_{o+}+\widetilde{\kappa}_{o-}\right)  \cdot\mathbf{B}~~\,,
\end{eqnarray}
where $\mathbf{E}$ and $\mathbf{B}$ respectively are the electric and magnetic field. To calculate the dispersion relation of the above Lagrangian, one must calculate its equations of motion. To this aim, it is better to use the initial Lagrangian. The first variation of \eqref{MGEh} with respect to $A_{\mu}$ leads:
\begin{equation}
\partial_{\alpha} F_{\mu}^{~\alpha} + (k_F)_{\mu\alpha\beta\gamma} \partial^{\alpha} F^{\beta\gamma}\,=\,0\,,
\end{equation}
and we also have the usual homogeneous Maxwell equation:
\begin{equation}
\partial_\mu *F^{\mu\nu}\,=\,0.
\end{equation}
For a plane electromagnetic wave with wave 4-vector $p^{\alpha}= (p^0, \vec{p})$, $F_{\mu\nu}= F_{\mu\nu}(p) e^{-i p_{\alpha} x^{\alpha}} $,   we then obtain the modified Ampere law \cite{Kostelecky:2001mb}:
\begin{equation}
M^{jk} E^{k} \equiv (-\delta^{jk} p^2 -p^{j}p^{k}- 2 (k_F)^{j\beta\gamma k} p_{\beta} p_{\gamma}) E^{k} \,=\, 0
\end{equation}
 The zero eigenvalues of  $M^{jk}$ identifies  the dispersion relation. To calculate the dispersion relation Let a prime coordinate be considered where in $\tilde{p}^{\alpha}= (p^0, 0,0, p^3)$. Assume that the anisotropy is not large. Then, in the prime coordinate, the zero eigenvalues at the leading order yields 
\begin{eqnarray}\label{BiF}
p_\pm^{0} &=& |p^3| (1 + \frac{k_{11}+ k_{22}}{2}\pm \sqrt{k_{12}^2 +\frac{(k_{11}-k_{22})^2}{4}}),\\
p_+^{0}& -& p_{-}^{0} \, = \,2 p^3 \sqrt{k_{12}^2 +\frac{(k_{11}-k_{22})^2}{4}}\,, \label{Bi1}
\end{eqnarray}
where 
\begin{equation}
k_{ij}= (\tilde{k}_F)_{i \alpha j\beta}\, \frac{\tilde{p}^{\mu} \tilde{p}^{\nu}}{|p_3|^2}\,,
\end{equation}
wherein $(\tilde{k}_F)_{i \alpha j\beta}$ represents the component of the $k_F$ tensor in the prime coordinate.   Of the above parameters,   $\tilde{\kappa}_{e+}$ and $\tilde{\kappa}_{o-}$ contribute to the  birefringence at the first order approximation \cite{Kostelecky:2001mb,Kostelecky:2002hh}. $\tilde{\kappa}_{e-}$ and $\tilde{\kappa}_{o+}$ do not contribute to the  birefringence at the first order approximation but they cause birefringence at the second order level \cite{Exirifard:2010xm}. \eqref{BiF} proves that $\tilde{\kappa}_{o+}$ and $\tilde{\kappa}_{o-}$ cause the one-way light speed anisotropy. $\tilde{\kappa}_{o+}$ has three components,  $\tilde{\kappa}_{o-}$ has five components. Therefore a general linear anisotropic media (of frequency dispersion relation) have eight parameters contributing to the one-way light speed anisotropy. 

Note that the Lorentz reciprocity lemma does not imply the isotropy of the one-way light speed when the effective Lagrangian for the light propagation includes terms in the form of multiplication of the electric and magnetic field strength. $\tilde{\kappa}_{o+}$ and  $\tilde{\kappa}_{o-}$ indeed lead to the one-way light speed anisotropy, as shown above.  

The parity can be broken by adding inhomogeneous impurity to the medium. The parity also can be broken by the molecular structure of the medium.  When the molecular structure of the medium breaks parity, there exists no theoretical argument that $\tilde{\kappa}_{o+}$ and  $\tilde{\kappa}_{o-}$ should vanish. They must  be measured. experimentally Examples of parity-odd crystals include the nematic liquid crystals and other crystals that have intrinsic  dipoles.

 \subsection{Electromagnetism of a none-linear medium}
The electromagnetic Lagrangian in a general non-linear material can be described by 
\begin{equation}
{\cal L}_{NL}\,=\,{\cal L} - \sum_{n=3}\frac{1}{4 n} \chi^{\mu_1\nu_1\cdots \mu_n\nu_n} \prod_{p=1}^{n} F_{\mu_p\nu_p}
\end{equation}
where $\cal L$ is the linear part of the Lagrangian. Let a strong constant field strength be turned on. So the field strength read
\begin{equation}
F = F_{ext} + \tilde{F}
\end{equation} 
where $\tilde{F}$ is the field strength of the propagating wave,  $|\tilde{F}| \ll |F_{ext}|$ and $F_{ext}=cte$. In this circumstance the non-linear Lagrangian can be approximated to
\begin{eqnarray}\label{20NL}
{\cal L} &=& -\frac{1}{4} \tilde{F}_{\mu\nu} \tilde{F}^{\mu\nu}\,-\frac{1}{4} (\tilde{k}_{F})_{\mu\nu\lambda\eta} \tilde{F}^{\lambda\eta} \tilde{F}^{\mu\nu} 
\end{eqnarray}
where 
\begin{equation}
(\tilde{k}_{F})_{\mu\nu\lambda\eta} = ({k}_{F})_{\mu\nu\lambda\eta} +   {\chi}^{\mu\nu\lambda\eta m n} (F_{ext})_{mn} + \cdots
\end{equation}
Notice that since $F_{ext}$ itself is a solution to the equation of motion then the terms  linear in $\tilde{F}$ in the action vanish. \eqref{20NL} beside its analogy with light propagation in linear anisotropic media, indicates that a sufficiently large one-way light speed anisotropy can be induced in a general non-linear crystal, in a crystal where the non-linear Lagrangian has terms  proportional to multiplication of some power of electric field and magnetic field. Perhaps this opens new windows to induce the one-way light speed anisotropy in non-linear media.\footnote{In theoretical physics, people are interested in the Born-Infled action:
\begin{equation*}
L = \beta^2 \sqrt{\det(\delta^{a}_b + \beta^{-2} F_a^{~b})}
\end{equation*}
This Lagrangian in the presence of  a constant electric and magnetic  field induces the one-way light speed anisotropy for the propagation of small electromagnetic field. 
}

\section{Optical diode}
\begin{figure}
\resizebox{0.75\columnwidth}{!}{%
  \includegraphics{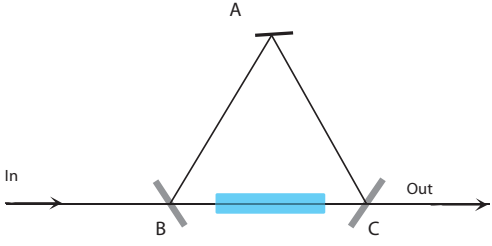}
}
\caption{Non-vacuum triangular resonator:  when the material exhibit one-way light speed anisotropy the resonator acts like a diode for its resonant frequencies.}
\label{fig:1}       
\end{figure}

In this section we illustrate how to construct optical diodes based on one-way light speed anisotropy. 
In so doing consider a triangular Fabry-Perot resonator \cite{Exirifard:2010xp} with a transparent material in one arm, fig. \ref{fig:1}. 
 Assume that the  transparent material is anisotropic such that the light speed in the material in the direction of $B$ to $C$ is not the same as that of  the direction of $C$ to $B$.     Also assume that the length of the  material is chosen such that the material does not change the polarization of the light ray should the anisotropic  material  rotates the polarization.  

\begin{figure}
\resizebox{0.75\columnwidth}{!}{%
  \includegraphics{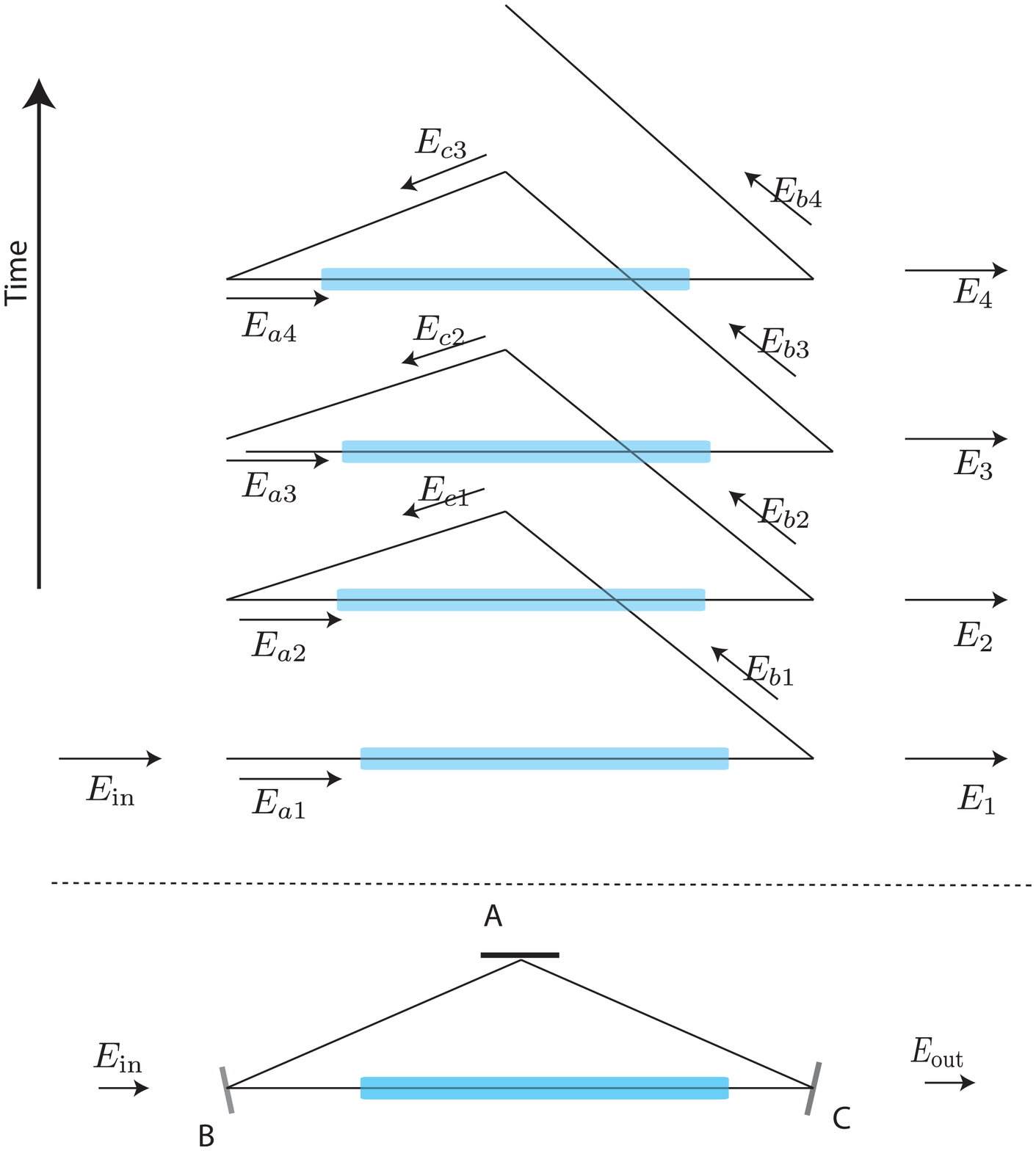}
}
\caption{Diagram of the derivation of the individual amplitudes}
\label{fig:2}       
\end{figure}

Now let us look at fig. \ref{fig:2}. The resonator consists of a perfect mirror at the corner $A$, two partial mirrors at the corners of $B$ and $C$. We send a single frequency light ray in the direction of $B$ to $C$ at the corner of $B$. For sake of simplicity of presenting the results,  assume that this light ray is polarized  such that its electric field is perpendicular to the surface of the triangle $ABC$.
Represent the electric field of the incoming light ray by $E_{\text{in}}$. The upper side of the fig. \ref{fig:2} depicts how the light ray is reflected back by the mirrors inside the resonator. Individual amplitudes inside the resonator reads:
\begin{eqnarray}
E_{a1} & =& \sqrt{t_b} E_{\text{in}}\\
E_{b1} & = & \sqrt{r_c} \sqrt{t_b} E_{\text{in}}  \\
E_{c1} & = & \sqrt{r_a r_c} \sqrt{t_b} E_{\text{in}}\\
E_{a2} & = & \sqrt{r_a r_b r_c} \sqrt{t_b} E_{\text{in}}
\end{eqnarray}
where $|r_a|, |r_b|$ and $|r_c|$ are the reflective index of the mirrors. They satisfy:
\begin{eqnarray}
|r_b| & = & 1 - |t_b| \\
|r_c| & = & 1 - |t_c| \\
|r_a| & =& 1.
\end{eqnarray}   
where $|t_b|$ and $|t_c|$ are the transitivity factors of the mirrors at the corners of $B$ and $C$. The outgoing light ray yields:
\begin{eqnarray}
E_{\text{out}} & =& E_1 + E_2 + E_3 + \cdots\,,\\
E_1 & =& \sqrt{t_b t_c} E_{\text{in}} \\
E_2 & =& \sqrt{r_a r_b r_c} \sqrt{t_b t_c} E_{\text{in}} e^{i\Delta\phi} \\
E_3 & =& r_a r_b r_c \sqrt{t_b t_c} E_{\text{in}} e^{2i\Delta\phi}\\
E_n & =& (\sqrt{r_a r_b r_c}  e^{i\Delta\phi})^n \sqrt{t_b t_c} E_{\text{in}}
\end{eqnarray}
where $\Delta \phi_+$ is the phase-shift of light rays after completing one circle inside the resonator. It reads:
\begin{equation}
\Delta \phi_+ \,=\, 2 \pi (\frac{L}{c} + (n_{+}-1) \frac{d}{c} ) \nu_+
\end{equation}
wherein $L$ is the perimeter of the triangle, and $n_{+}$ is the reflective index of the  material in the direction of $B$ to $C$, $d$ stands for its length, and $c$ is the light speed in the vacuum, and $\nu_+$ is the frequency of the light. So the out coming light ray reads:
\begin{equation}
E_{\text{out}} \, =\, \frac{\sqrt{t_b t_c}} {1 - \sqrt{r_a r_b r_c} e^{i\Delta \phi_+}} E_{\text{in}}
\end{equation}
Defining 
\begin{eqnarray}
r_a & =& e^{i\delta_a} \\
r_b & =& |r_b| e^{i\delta_a} \\
r_c & =& |r_c| e^{i\delta_a}
\end{eqnarray}
the out-coming ray is re-expressed by 
\begin{equation}
E_{\text{out}} \, =\, \frac{\sqrt{t_b t_c}} {1 - \sqrt{|r_b r_c|} e^{i\Delta \phi_++ i (\delta_a + \delta_b + \delta_c)}} E_{\text{in}}
\end{equation}
For the frequencies that yield
\begin{equation}\label{anticlockwise_resonator}
\Delta \phi_++ \delta_a + \delta_b + \delta_c \,=\, 2 \pi n
\end{equation}
where $n$ is a natural number ($n$ labels resonant frequencies)  if we choose the same partial mirrors at $B$ and $C$, 
\begin{eqnarray}
t_b &=& t_c \, \equiv\, t \\
r_b & =& r_c\, \equiv\, r 
\end{eqnarray}
then  the incoming ray is completely  transmitted by the resonator. 

The resonant frequency for lights moving clockwise in the triangular resonator differs from \eqref{anticlockwise_resonator} because an asymmetric  material is used in the resonator. The clockwise resonant frequencies read:
\begin{eqnarray}\label{clockwise_resonator}
\Delta \phi_- \,=\, 2\pi(\frac{L}{c} + (n_{-}-1) \frac{d}{c}) \nu_-\\ 
\Delta \phi_-+ \delta_a + \delta_b + \delta_c \,=\, 2 \pi n
\end{eqnarray}
where $n_-$ is the reflective index of the  material in the direction of $C$ to $B$ , and $\nu_-$ is the frequency of the light. 

Consider a resonant frequency. In one hand,  the sharpness of this resonant frequency ($\frac{\Delta \nu}{\nu}$) is proportional to $\frac{t}{r}$:
\begin{equation}
\left(\frac{\Delta \nu}{\nu}\right)_{\text{Res.}} \propto \frac{t}{r} 
\end{equation}
Nowadays we can make this resonant frequency  as sharp as $10^{-10}$  in the  commercial equipments and even sharper  for  the scientific purposes. 

\begin{figure}
\resizebox{0.75\columnwidth}{!}{%
  \includegraphics{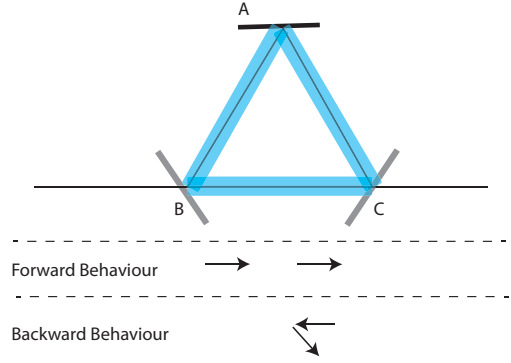}
}
\caption{The response of the optical diode for forward and backward currents.}
\label{fig:3}       
\end{figure}

In the other hand, the difference between the same modes of the clockwise \eqref{clockwise_resonator} and anti-clockwise resonant \eqref{anticlockwise_resonator} frequencies reads:
\begin{equation}\label{rf}
2\frac{\nu_+ - \nu_- }{\nu_+ + \nu_-}
\,\approx \,\frac{  d  (n_- - n_+)} {L + d (\frac{n_+ + n_-}{2}-1)} 
\end{equation}
where we have approximated 
\begin{equation}
n_+- n_-\ll \frac{1}{2}(n_- + n_+),.
\end{equation}
When the resonant frequencies are sharp enough then the clockwise and anti-clockwise resonant frequencies are not the same. This phenomenon happens when
\begin{equation}
2\frac{\nu_+ - \nu_- }{\nu_+ + \nu_-} > \left(\frac{\Delta \nu}{\nu}\right)_{\text{Res.}}
\end{equation}
or equivalently
\begin{equation}\label{above}
\frac{t}{r}< \frac{2|n_+ - n_-|}{\frac{L}{d}-1 + \frac{n_+ + n_-}{2}}
\end{equation}
The right hand side of the above inequality reaches its maximum value when $L=d$. This suggests  to appropriately fill all the edges of the resonator with the anisotropic material in order to enhance the difference between the clockwise and anti-clockwise frequencies; fig. \ref{fig:3} . If we do so then \eqref{above}  simplifies to
\begin{equation}\label{final}
\frac{t}{r} < \frac{2 |n_+ - n_-|}{n_+ + n_-}
\end{equation}
Slight  light absorption in the anisotropic material does not significantly alter the above relation. This is due to the fact that  by choosing sufficiently a small triangle we can make the absorption  sufficiently small. In practice it is possible to make $\frac{t}{l}$ sufficiently small, smaller than $10^{-10}$.  So when a one-way light speed anisotropic transparent material, possessing  an anisotropy larger than $\frac{|n_+ - n_-|}{n_+ + n_-} > 10^{-10}$,  is found then \eqref{final} is met  and the triangular resonator  serving as an optical diode can be constructed. Fig. \ref{fig:3} depicts how this diode  functions: the anti-clockwise resonant frequencies are transmitted to the other side. But if we send back these frequencies from the other side of the resonator, in the directions that they come out of the resonator, they shall be reflected back. This establishes a perfect optical diode.

\section{Measuring the one-way light speed anisotropy}
In the first section we review the light propagation in a general anisotropic media. In the second section we show how to harvest the one-way light speed anisotropy to construct new optical diodes. The experimental community, however so far, has not been interested to measure the one-way light speed anisotropy in any medium that breaks the parity.  We urge the experimental groups to measure the one-way light speed anisotropy in the linear crystals that break the parity. Crystals possessing intrinsic electric dipole breaks parity. So the nematic liquid crystals have a chance to exhibit the one-way light speed anisotropy. 

\begin{figure}
\resizebox{0.75\columnwidth}{!}{%
  \includegraphics{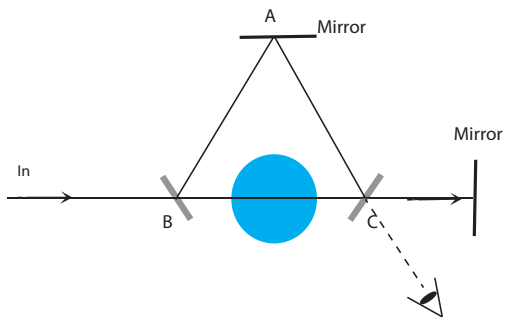}
}
\caption{A simple setup to observe and measure the one-way light speed anisotropy. Unpolarized white light is emitted to the resonator, the resonant frequencies are reflected back into the resonator.  The part that does not go into the resonator is observed.}
\label{fig:4}       
\end{figure}

The one-way light speed anisotropy can be measured by a Mach-Zhender interferometer.   Another  setup is based on the optical-diode,  depicted in Fig. \ref{fig:4}: it suffices to make a triangular resonator,  place a piece of the anisotropic material in the resonator and reflects back the outgoing ray into the resonator. Now rotates the material. When the outgoing ray does not enter into the resonator then the one-way anisotropy is observed.  Note that this setup is robust to the environmental noises. It is not needed to perform the experiment in low temperature, vacuum or on a very highly stable optical table. Also the beat frequency between  the clockwise and anti-clockwise resonate  frequencies is another physical quantity that encodes the one-way light speed anisotropy in the crystals. Measuring this beat frequency is  simple too.  

  In the previous section we show that a tiny one-way asymmetry, an asymmetry  of orders of parts in ten billions leads to the construction of new optical diode.  So we urge the experimental groups,  using  their desired setups, to measure the one-way light speed isotropy or anisotropy with the precision of parts in ten billions in the nematic liquid crystals   as well as other anisotropic parity-odd materials, or in none-linear media in the presence of constant field strengths.   Measuring the one-way light speed anisotropy is simple, and should we observe a one-way anisotropy larger than parts in ten billions the reward is astonishing: new optical diodes are constructed. 

\section{Conclusions and outlook}
We have used the SME electrodynamics to address  the light propagation in a general anisotropic  media. We have realized that when parity is broken by the molecular structure of the linear material, or constant field strength is applied on a general non-linear material, then the one-way light speed can be anisotropic. In these cases, the anisotropy or isotropy of the one-way light speed must be empirically measured. 

We have shown how to utilize a tiny one-way light speed anisotropy, as small as  parts in ten billions, to construct new perfect optical diodes. So we have urged the experimentalists to measure the intrinsic one-way light speed in various parity-odd linear crystals, or the induced one-way light speed anisotropy in none-linear materials with this precision.

\section*{Aknowledgements}
I thank M. Khorrami, Z. Fakhrai, A. Moradi and J. Niemela for discussions and correspondences. This work was partially supported by the school of physics, IPM. 
%

\begin{thebibliography}{}
%
%


\bibitem{NatureDiode}
Zheng Wang, Yidong Chong, J. D. Joannopoulos, Marin Soljacic,  \textit{Observation of unidirectional backscattering-immune topological electromagnetic states},
Nature {\bf 461} (2009).


\bibitem{ScienceDiode}
Liang Feng, Maurice Ayache, Jingqing Huang, Ye-Long Xu, Ming-Hui Lu, Yan-Feng Chen, Yeshaiahu Fainman, Axel Scherer,
\textit{Nonreciprocal Light Propagation in a Silicon Photonic Circuit},
Science {\bf  333} (2011) 729.

\bibitem{NonLinear1}
K. Gallo, G. Assanto, K. R. Parameswaran, M. M. Fejer,
\textit{ All-optical diode in a periodically poled lithium niobate waveguide},
Appl. Phys. Lett. {\bf 79} (2011) 314,

\bibitem{NonLinear2}
Z. Yu, S. Fan, 
\textit{Complete optical isolation created by indirect interband photonic transitions.}
Nat. Photonics {\bf 3} (2009) 91. 


\bibitem{N2}
Wei-Min Ye, Xiao-Dong Yuan, and Chun Zeng, \textit{Unidirectional transmission realized by two nonparallel gratings made of isotropic media,} Opt. Lett. \textbf{36}, 2842-2844 (2011) 


\bibitem{M1}
V. A. Fedotov, P. L. Mladyonov, S. L. Prosvirnin, A. V. Rogacheva, Y. Chen, and N. I. Zheludev,
\textit{Asymmetric Propagation of Electromagnetic Waves through a Planar Chiral Structure},
Phys. Rev. Lett. \textbf{97} (2006) 167401. 


\bibitem{M2}
C. Menzel, C. Helgert, C. Rockstuhl, E.-B. Kley, A. Tunnermann, T. Pertsch, and F. Lederer, \textit{Asymmetric Transmission of Linearly Polarized Light at Optical Metamaterials}, Phys. Rev. Lett. \textbf{104} (2010) 253902.


\bibitem{M3}
Ilya V. Shadrivov, Vassili A. Fedotov, David A. Powell, Yuri S. Kivshar and Nikolay I. Zheludev, 
\textit{Electromagnetic wave analogue of an electronic diode}
New J. Phys. \textbf{13} (2011) 033025


\bibitem{M4}
Wan-xia Huang, Yi Zhang, Xia-mei Tang, Li-Sha Cai, Jun-wei Zhao, Lin Zhou, Qian-jin Wang, Cheng-ping Huang, and Yong-yuan Zhu, \textit{Optical properties of a planar metamaterial with chiral symmetry breaking,} Opt. Lett. \textbf{36} (2011) 3359-3361.

 
\bibitem{M5}
C. Menzel, C. Helgert, C. Rockstuhl, E.-B. Kley, A. TŸnnermann, T. Pertsch, and F. Lederer,
\textit{Asymmetric Transmission of Linearly Polarized Light at Optical Metamaterials}, 
Phys. Rev. Lett. \textbf{104}, 253902 (2010). 




\bibitem{Kostelecky:2001mb}
  V.~A.~Kostelecky and M.~Mewes,
 \textit{Cosmological constraints on Lorentz violation in electrodynamics},
  Phys.\ Rev.\ Lett.\  {\bf 87} (2001) 251304 [arXiv:hep-ph/0111026].
  



\bibitem{Kostelecky:2002hh}
  V.~A.~Kostelecky and M.~Mewes,
  \textit{Signals for Lorentz violation in electrodynamics},
  Phys.\ Rev.\  D {\bf 66} (2002) 056005, [arXiv:hep-ph/0205211].
  

\bibitem{Kostelecky:2008ts}
  V.~A.~Kostelecky, N.~Russell,
  \textit{Data Tables for Lorentz and CPT Violation},
  Rev.\ Mod.\ Phys.\  {\bf 83}, 11 (2011),
  [arXiv:0801.0287 [hep-ph]].

\bibitem{Exirifard:2010xm}
  Q.~Exirifard,
  \textit{Cosmological birefringent constraints on light,}
  Phys.\ Lett.\  {\bf B699}, 1-4 (2011).
  [arXiv:1010.2054 [gr-qc]].

\bibitem{Exirifard:2010xp}
  Q.~Exirifard, \textit{Triangular Fabry-Perot resonator:
Direct measurement of the parity-odd terms of the photon-sector of the SME}, [arXiv:1010.2057 [gr-qc]].


\end{thebibliography}
%

\end{document}